\documentclass[a4paper,twoside,12pt]{article} 

\input{sp.sty} 

\def\spvol{1}									
\def\spissue{1}									
\def\sppage{1}									
\def\spmonth{July}								
\def\spyear{2022}								

\def\spshort{The eclipse timings of RR Lyncis}	
\def\spshauth{C. Lloyd}							

\begin{document}
	
\setcounter{page}{\sppage}

\SPbigheader{\spshort}{\spshauth}{\spmonth~\spyear}{\spvol}

\SPheader{\spshort}{\spshauth}{\spmonth~\spyear}{\spvol}{\spissue}

\SPtitle{The eclipse timings of RR Lyncis}

\SPauth{
	Christopher Lloyd 
}


	\SPinstone{School of Mathematical and Physical Sciences, University of Sussex}

\SPabstract{
	The times of minima of the bright, eccentric eclipsing binary RR Lyncis are re-investigated.  From the TESS data it is found that significant differences in the eclipse timings, particularly for the primary minima, are probably due to small, irregular changes in the eclipse profile due to pulsations, rather than light-curve asymmetries.
}

\section*{Introduction}

RR Lyncis is a naked-eye detached eclipsing binary with a period near 9\fday95, having eclipses of  $\sim 0\fmm3$ depth and a small eccentricity.  
A major reinvestigation of the system using radial velocity and new TESS data together with previous  photometry has recently been published by
\cite{2021Obs...141..282S}. 
The present interest in the star concerns the possibility of a third body in the system as suggested by \cite{2002ARep...46..119K}. The purpose of this paper is to provide supporting material to the reinvestigation of the third-body orbit by \cite{lloydinpress}. 

Some of the previously published eclipse timings do not include the heliocentric corrections, and others are averages of times from different filters, which it is felt could conceal uncertainties in individual minima. So, with that in mind, where possible the original, individual light-curves  have been remeasured and some new data are considered.

Two methods have been employed to fit the times of minima, depending on which is most appropriate for the data. The first is the standard 
Kwee-van Woerden method (KvW) \citep{1956BAN....12..327K}, and the second is a simple Gaussian fit to the minimum and part of the constant section of the light-curve. Obviously this has limited application, but for partial eclipses in detached systems, particularly when the data are noisy or sparse, it can be very useful. Both approaches may be used on time-series data or on folded data to derive composite timings, and for relatively well-covered minima they are usually consistent, with similar uncertainties. In some cases one is preferred for whatever reason. The different data sets are discussed below. The phase diagrams have all been constructed using $P=9.94507108$ days and the derived composite timings are not sensitive to this value.

\section*{Huffer}

\begin{figure}[!t]
	\centering
	\includegraphics[width=.49\textwidth]{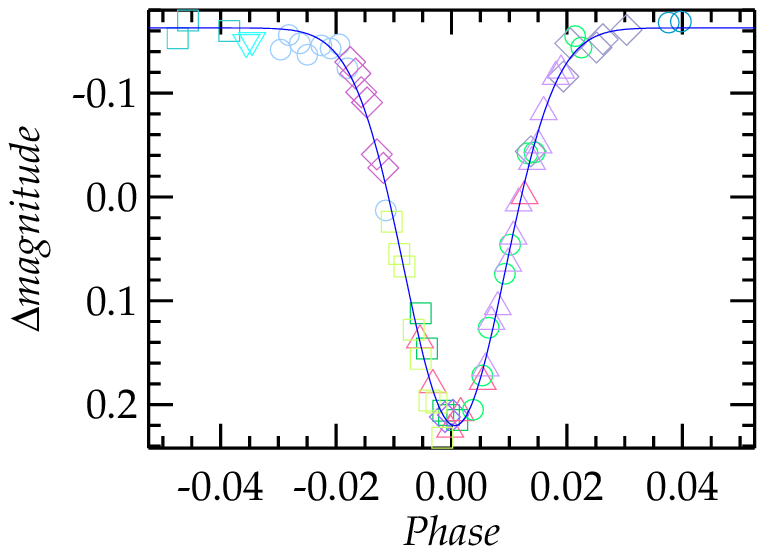}
	\includegraphics[width=.49\textwidth]{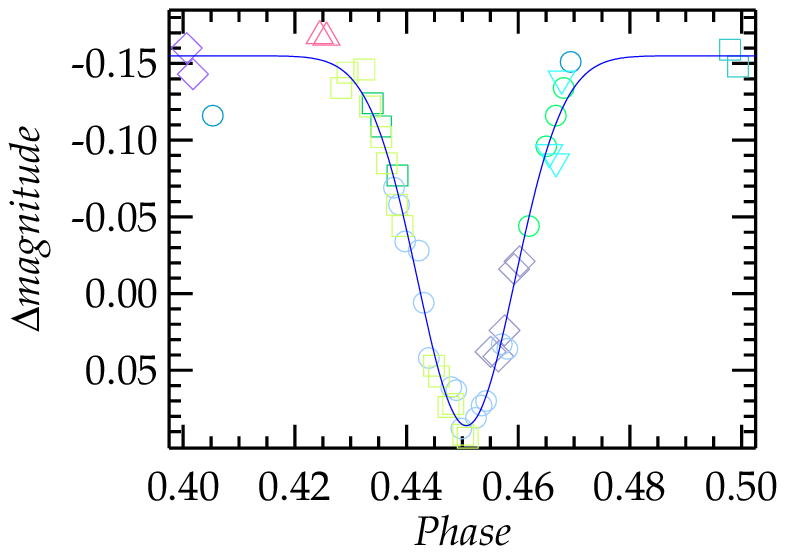}
	\caption{Phase diagram of Huffer's data showing detail of the primary and secondary minima. The lines show the Gaussian fit to the data, for illustrative purposes only. The different symbols identify different runs.}
	\label{fig:huffer}
\end{figure}

Huffer observed RR~Lyn with an early photoelectric photometer built by Stebbins and collaborators 
\citep{1928PWasO..15D...2S} which had a panchromatic blue response peaking near 4600~\AA, and in modern parlance probably most resembles Johnson $B$-band. A total 154 measurements were made between 1923 December and 1929 January, but these were mostly in small groups of often only two observations, up to a maximum of 14 on two nights. Although some runs do cover the minima there are not enough observations to derive reliable timings so it is necessary to fold the data and derive composite measures.
Huffer gave an ephemeris and the offset of the secondary minimum which have been passed down, but these do not contain the heliocentric correction. For completeness the times have been redetermined from the data, taken from Table~II of 
\cite{1928PWasO..15..199H}, and shown in Figure~\ref{fig:huffer}. Both the KvW and Gaussian methods have been used here to calculate the times of minima and provide similar values, well within the errors, but the KvW values have been adopted as the method is more general. 
Huffer aligned his ephemeris on a run covering primary minimum that occurred towards the end of his data, and the same epoch has been used here for consistency, but the body of the data were acquired three or more years earlier. The timings are collected with the others in Table~\ref{tab:tom}. 

\section*{Botsula}

\begin{figure}[!b]
	\centering
	\includegraphics[width=.49\textwidth]{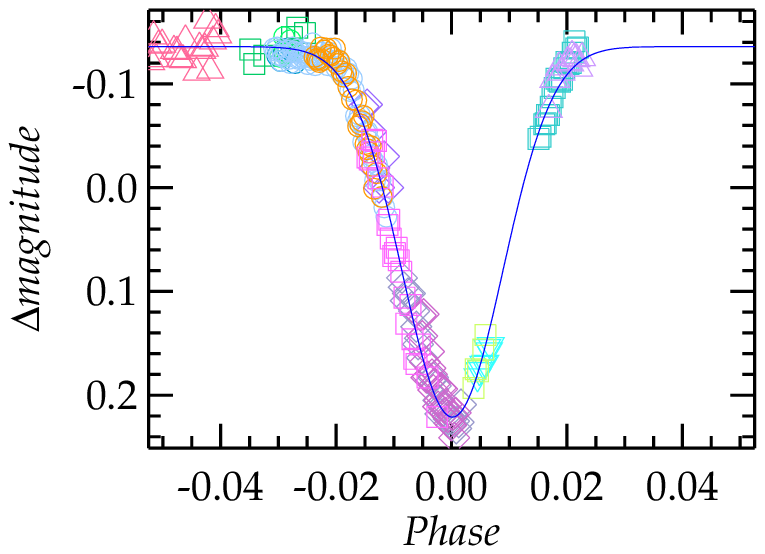}
	\includegraphics[width=.49\textwidth]{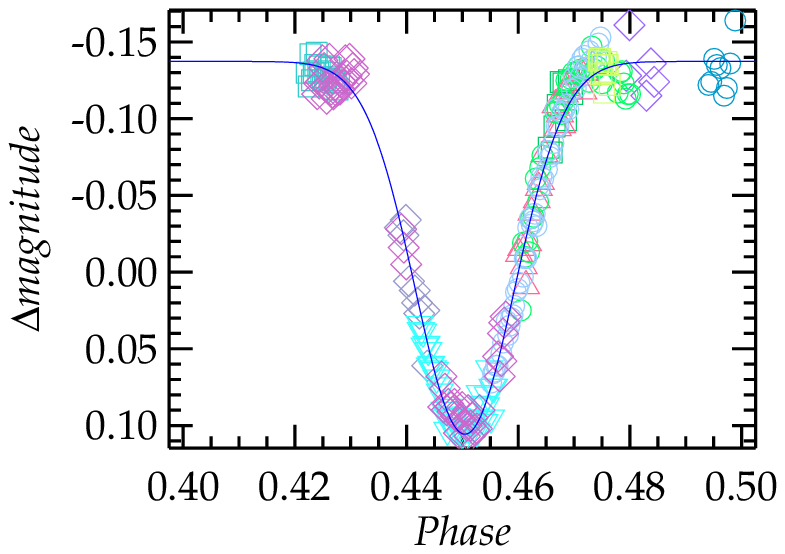}
	\caption{Phase diagram of Botsula's data showing detail of the primary and secondary minima. The lines show the Gaussian fit to the data, which was used for the timing in this case. The different symbols identify different runs. The poor coverage of the primary egress is clear.}
	\label{fig:botsula}
\end{figure}

In a similar manner \cite{1960Botsula} made an extensive of observations between 1949 September and 1956 April that were discussed in detail some years later \citep{1968SvA....11.1000B}. These runs are better populated with up to 66 points, but again, none of them cover enough of a minimum to extract a timing. Even the combined data have some significant gaps in coverage, particularly the primary minimum. Botsula gives two timings but these again do not include the heliocentric correction. The timings were  remeasured using both methods and these agree within the errors, but given that the primary minimum is poorly covered the values from the Gaussian fit have been adopted, and these are given in Table~\ref{tab:tom}. Botsula's data have also been re-analysed by \cite{2002ARep...46..119K} and while their primary minimum is consistent with the one derived here the secondary timing differs by 0\fday0015, which is a little outside the combined errors. Botsula's minima are shown in Figure~\ref{fig:botsula}.

\section*{Linnell}

\cite{1966AJ.....71..458L} made an extensive series of $UBV$ observations on 16 nights between 1960 November and 1963 January. Most of these were also in short runs but three night cover significant parts of the eclipses, although the egress of the secondary eclipse is not well observed. Linnell combined his data to give one primary and secondary timing, but the individual filters are analysed separately here. The primary minima are well covered and provide consistent timings. The secondary on the other hand has limited coverage and is interrupted, presumably by cloud. The $U$ and $B$ secondary timings are very discordant, but the $V$ timing is consistent with other data, although it has a large uncertainty.

Linnell also examined the period behaviour of RR~Lyn with the data available at the time and found that the times given by \cite{1959AbaOB..24...13M} 
were seriously discordant with Botsula's ephemeris and required correction to bring them into better alignment. These times of minima were not used by \cite{2002ARep...46..119K}.

\begin{figure}[!h]
	\centering
	\includegraphics[width=.49\textwidth]{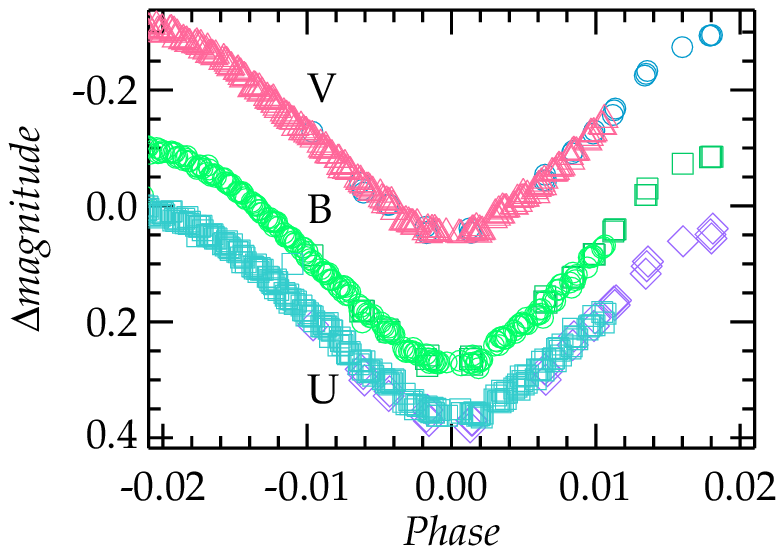}
	\includegraphics[width=.49\textwidth]{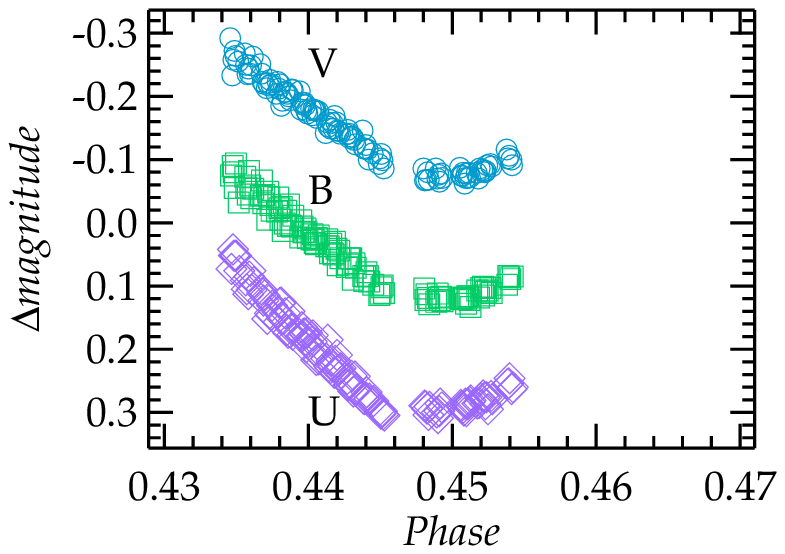}
	\caption{Phased time-series plot of Linnell's data showing detail of the two primary minima in three bands, and similarly for the single secondary minimum. The different symbols identify different runs. The poor coverage of the secondary minimum is clear and leads to compromised timings.}
	\label{fig:linnell}
\end{figure}

\section*{Khaliullin, Khaliullina, \& Krylov}

\begin{figure}[!t]
	\centering
	\includegraphics[width=.49\textwidth]{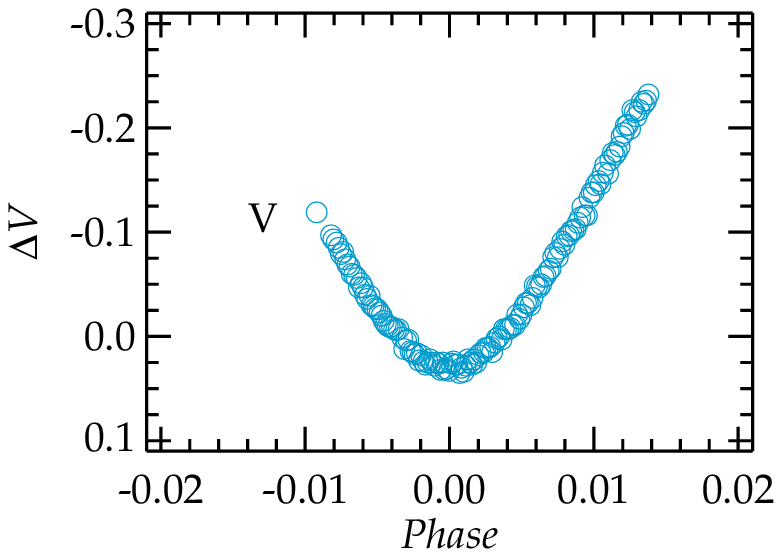}
	\includegraphics[width=.49\textwidth]{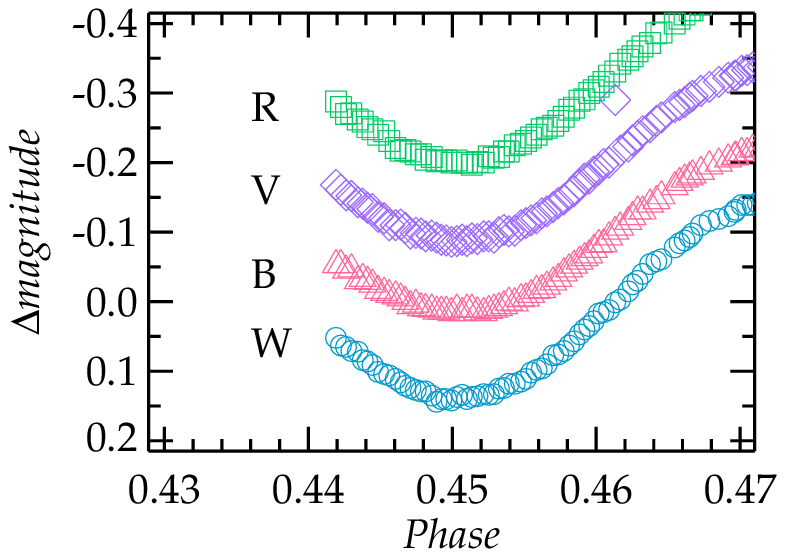}
	\caption{Phased time-series data of Khaliullin et al. showing detail of the primary and secondary minima in various bands.}
	\label{fig:kkk}
\end{figure}

High-precision, but rather limited multi-colour observations are reported by \cite{2001ARep...45..888K} that were made in 1980 December, 1981 January and 1982 January. One run was made through the primary minimum in $V$ with another through the secondary in $W, B, V$ and $R$, and other short sets of multi-colour observations were made at other times during the constant part of the light-curve. The phased time-series runs are shown in Figure~\ref{fig:kkk} and five independent timings are derived using the KvW method, which are included in Table~\ref{tab:tom}. The symmetrical part of the secondary eclipse, which is what KvW requires, is not very deep, barely 0\fmm1 in all colours, and this limits the consistency, with the $B$-band timing being the slightly removed from the others.

Khaliullin \etal\ used a photometric model of the system to derive their times of minima rather then direct measurement, so there is potential for some disagreement. However, their time of primary minimum is later by only 0\fday00031, and the secondary is later by 0\fday00018 than the mean of the independent timings, both of which are well within the combined errors. 

\begin{figure}[!b]
	\centering
	\includegraphics[width=.49\textwidth]{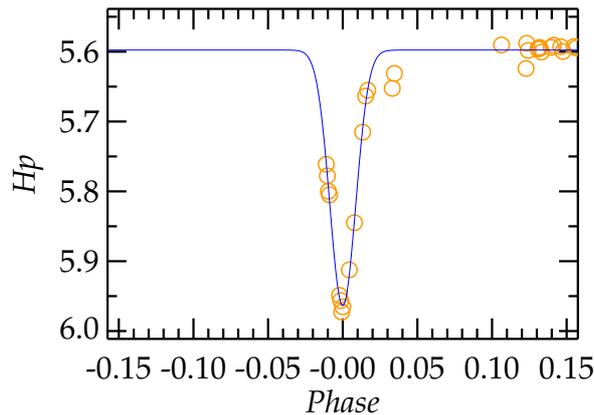}
	\caption{Phase diagram of the Hipparcos data showing detail of the primary minimum and the Gaussian fit.}
	\label{fig:hipp}
\end{figure}

\section*{Hipparcos}

The Hipparcos mission provides nominally high-precision observations of RR~Lyn but unfortunately these are relatively few, and only the primary eclipse is covered, but not well. The phase diagram is shown in Figure~\ref{fig:hipp} together with the Gaussian fit that yields a low-precision timing.

\section*{MASCARA}

\begin{figure}[!h]
	\centering
	\includegraphics[width=.49\textwidth]{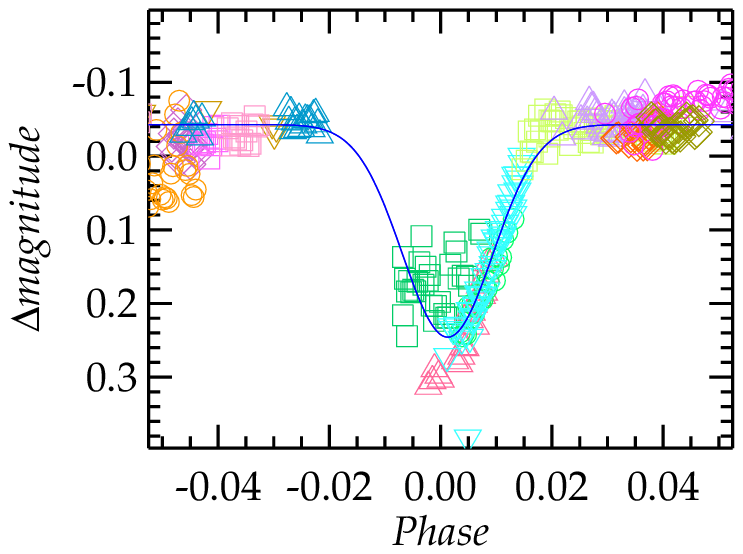}
	\includegraphics[width=.49\textwidth]{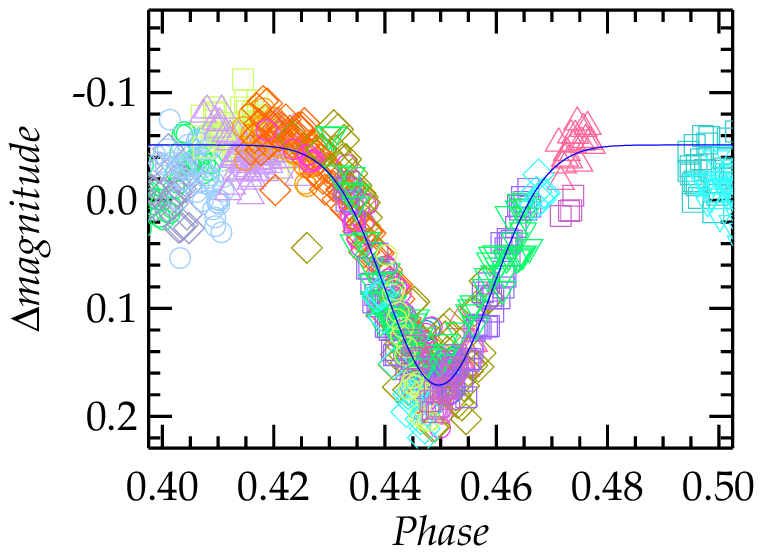}
	\caption{Phase diagram of the MASCARA data, with the Gaussian fits shown. The primary timing is unreliable for obvious reasons, and the secondary provides only a low-precision value.}
	\label{fig:mascara}
\end{figure}

More recent data are also available from the Multi-site All-Sky CAmeRA (MASCARA) project \citep{2018A&A...617A..32B}, which can produce photometry on bright stars ($4 < V < 8.5$) to a precision of $\sim 0\fmm03$. Unfortunately, the data as presented still contain significant systematics and lower-precision data. The phase diagrams showing the detail around the minima are given in Figure~\ref{fig:mascara}. Given the density of the data and the scatter, the Gaussian fit has been used, but the primary is poorly covered and yields a very discordant timing. The secondary eclipse does provide a timing, but it is of limited value in the contest of RR~Lyn, and require data to be sigma-clipped from the solution.

\section*{TESS}

\begin{figure}[p]
	\centering
	\includegraphics[width=.49\textwidth]{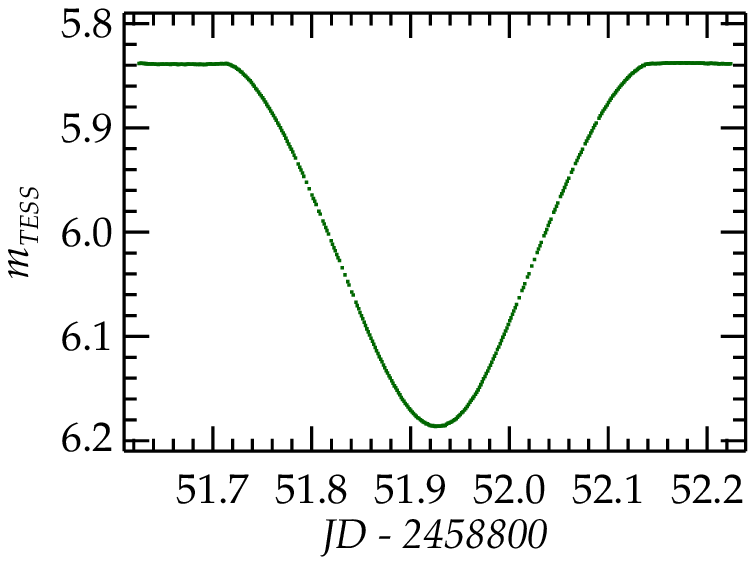}
	\includegraphics[width=.49\textwidth]{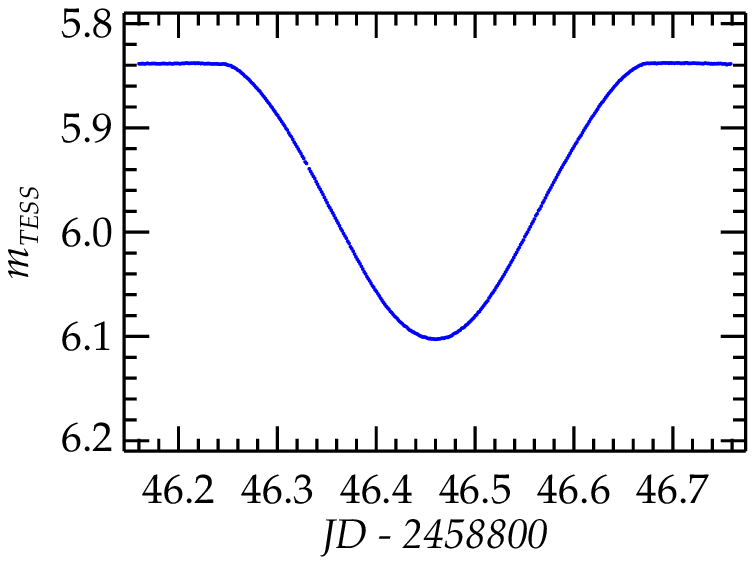}
	\caption{TESS time-series data showing a primary minimum (left) and a secondary minimum (right). At this scale similar eclipses are indistinguishable.}
	\label{fig:tess}
\end{figure}

\begin{figure}[p]
	\centering
	\includegraphics[width=.49\textwidth]{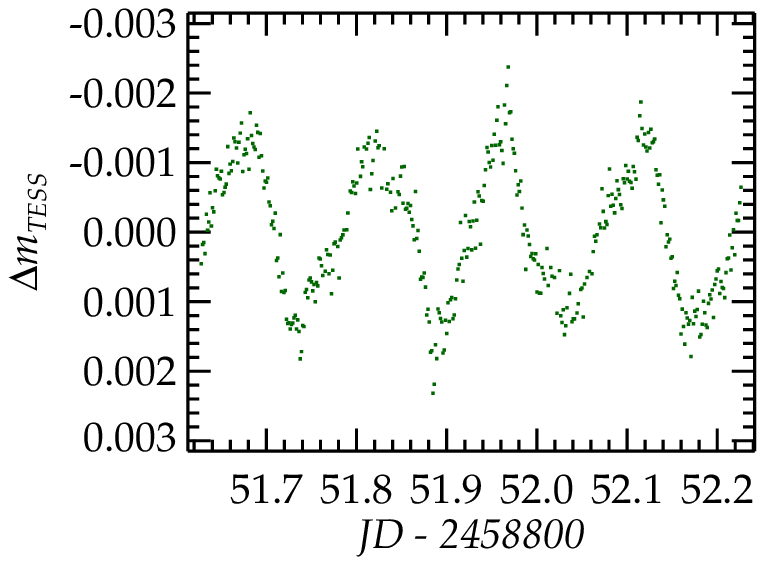}
	\includegraphics[width=.49\textwidth]{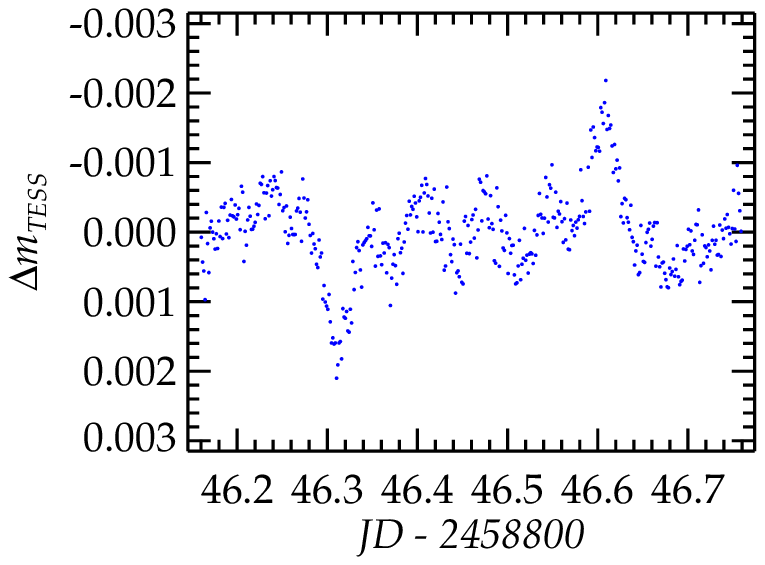}
	\includegraphics[width=.49\textwidth]{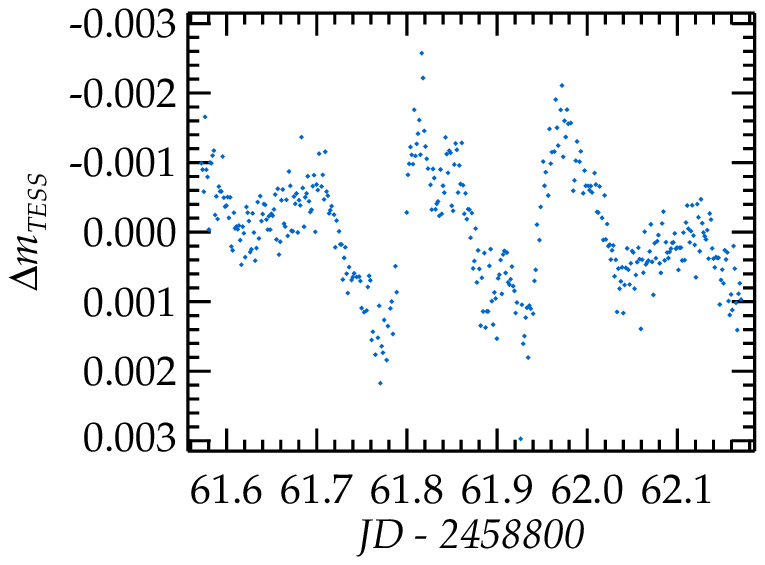}
	\includegraphics[width=.49\textwidth]{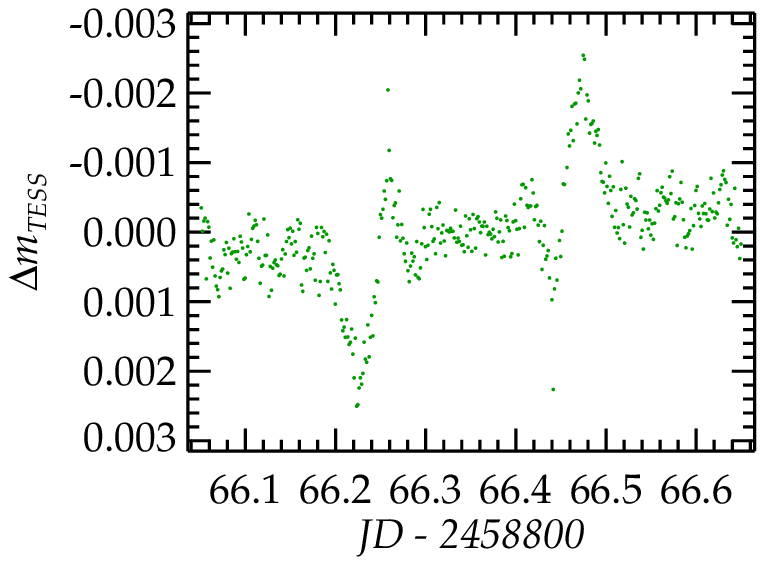}
	\caption{Magnitude-difference plots for the four TESS minima showing the reverse light-curve relative to the forward one for the primary minima (left) and the secondary minima (right). The reflection point used is the timing from the $m_{TESS}=6.00$ KvW measurement as this is in the middle of the range, but the distribution is not sensitive the the value used. The primary minima are very different and probably dominated by chance combinations of the pulsation variations, while for the secondary minima these appear suppressed, but there may be some evidence of consistent distortion in the two eclipses.} 
	\label{fig:reverse}
\end{figure}

The primary component of RR~Lyn has been known as an Am star for many years 
but the TESS data also reveal that the system shows at least 35 low-amplitude pulsations in the $\delta$~Scuti and $\gamma$ Doradus range
\citep[see][]{2021Obs...141..282S}. 
The amplitudes are very low, $< 0.3$~mmag, and the excursions from the mean light-curve are $< 3$~mmag so they will have little impact on the ground-based light-curves. Two of the TESS minima are shown in Figure~\ref{fig:tess} by way of illustration.
Southworth showed (in his Figure~2) that the dispersion around his photometric model is typically $\sim 1$~mmag for the primary eclipse and significantly less during the secondary eclipse, leading to the conclusion that the pulsations occur on the secondary component. A similar result can be seen in Figure~\ref{fig:reverse} where the magnitude difference between the light-curve and its reflection are shown for all four TESS eclipses. The primary minima while different, show rapid variations peaking at 1-2~mmag, however, there is no obvious trend that would point to any clear asymmetry in the eclipses. The variation in the secondary minima is generally much smaller but there are two consistent features corresponding to the steepest descent of the eclipse that might point to asymmetry.

RR~Lyn has long been a subject of investigations of possible apsidal motion but although these have not been found they have permeated the discussions, and Khaliullin \etal\ implicitly assumed they were present. Examination of the TESS data by
\cite{2021A&A...649A..64B} found no indication of apsidal motion but did identify asymmetries in the shape of the minima. In eccentric eclipsing binaries these distortions are anticipated but their scale, and effect on the times of minima, is not necessarily clear. To examine this the KvW method has been used to derive timings from decreasing portions of the TESS light-curve. The constant part lies at $m_{TESS}=5.84$ and the KvW method has been applied to the minimum below $m_{TESS}=5.90$, and subsequently fainter levels.

\begin{figure}[!b]
	\centering
	\includegraphics[width=.75\textwidth]{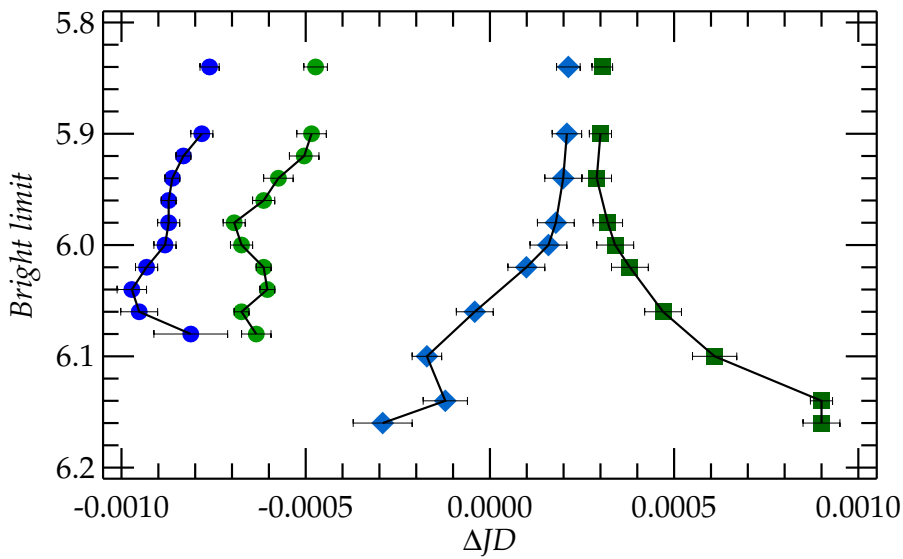}
	\caption{The times of minima from \cite{2021A&A...649A..64B} relative to Southworth's values  are shown towards the top of the plot. The connected points correspond to the different KvW samples from each TESS eclipse. The ordinate gives the bright limit of the data used in the KvW sample and $\Delta JD$ gives the difference relative to the time of primary or secondary minimum from Southworth. The migration of the measured time of minimum for the primary (squares and diamonds) and secondary (circles) eclipses are clear as progressively deeper parts of the eclipses are used. The times from Baroch \etal\ are derived from the entirety of the eclipse. The obvious offset of the secondary minima is due to the relatively low-precision value of $\phi_2$ given by Southworth, which was never intended to be used as a timing of secondary minimum. Optimizing $\phi_2$ for the TESS data will improve the alignment of the primary and secondary timings but cannot bring them within the measured uncertainties.}
	\label{fig:asymmetry}
\end{figure}
 
\cite{2021Obs...141..282S} 
used a photometric model to fit the TESS data and derived the time of primary minimum together with a measure of the phase difference of the secondary, $\phi_2$, and these have been used as the reference points for this discussion.
Figure~\ref{fig:asymmetry} shows the time difference for each of the samples from the four TESS minima, as indicated by the bright limit used in the measurement. The primary minima show very different behaviour but both show substantial migration of the derived timing with decreasing depth of the sample amounting to $\pm0\fday0005$. The uncertainties are typically a factor of ten smaller. The secondary timings are much more similar, and have a smaller range of 0\fday0002, but from a sample of two, all that can be said is that they are not inconsistent. The difference between the two primary eclipses suggests that the migration is not due to any asymmetry in the eclipse profile, as that would be repeatable, but is more likely to be due to low-level variation from the pulsations, and this is consistent with the interpretation of Figure~\ref{fig:reverse}. The times measured by \cite{2021A&A...649A..64B} are shown as the isolated points towards the top of the plot and all these are close to measurements based on the major part of the eclipses, which presumably suppresses the timing variations. 

\begin{figure}[!b]
	\centering
	\includegraphics[width=.75\textwidth]{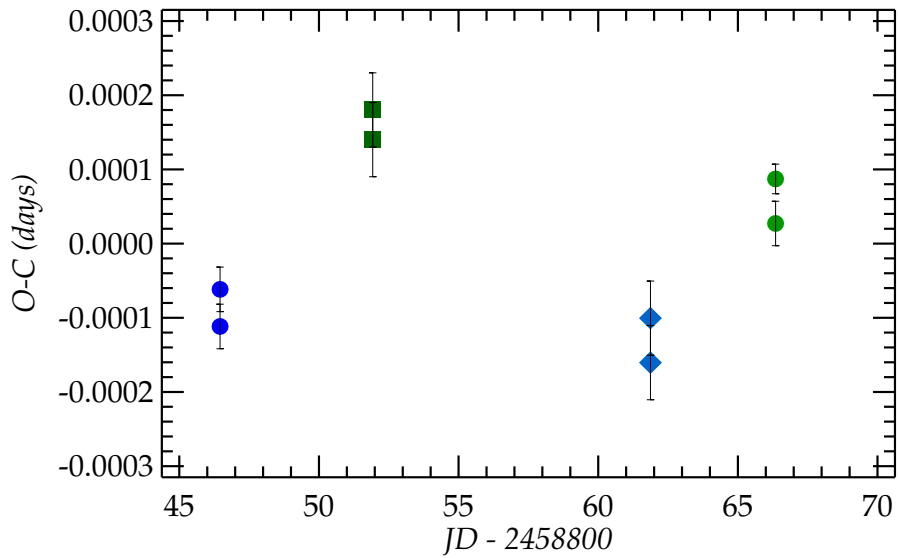}
	\caption{The O-C diagram of the KvW timings with bright limits of $m_{TESS}=6.00$ and 6.02 from Figure~\ref{fig:asymmetry}. These represent the most consistent timings. The residuals are from solution with the optimized period and  $\phi_2$.}
	\label{fig:residuals}
\end{figure}

\begin{table}[t]
	\caption{\em Times of minimum \label{tab:tom}}
	\centering
	\addtolength{\tabcolsep}{-3pt} 
	\begin{tabular}{ccccl}
		{\em HJD} & {\em Error (d)} & {\em Min.} & {\em Band} & {\em Data set} \\
		\addlinespace[3pt] 
		2425615.4956~  & 0.0013~  & 1 &  $(B)$ & Huffer \\ 
		2425619.9741~  & 0.0016~  & 2 &  $(B)$ & Huffer \\ 
		2434675.4581~  & 0.0007~  & 1 &  $(B)$ & Botsula \\ 
		2434679.9364~  & 0.0008~  & 2 &  $(B)$ & Botsula \\ 
		2437698.7597~  & 0.0005~  & 1 &  $B$ & Linnell \\ 
		2437698.7598~  & 0.0007~  & 1 &  $V$ & Linnell \\ 
		2437698.7604~  & 0.0010~  & 1 &  $U$ & Linnell \\ 
		2437941.9201~  & 0.0015~  & 2 &  $V$ & Linnell \\ 
		2438046.83761  & 0.00033  & 1 &  $B$ & Linnell \\ 
		2438046.83765  & 0.00028  & 1 &  $V$ & Linnell \\ 
		2438046.83992  & 0.00029  & 1 &  $U$ & Linnell \\ 
		2444595.17158  & 0.00047  & 2 &  $W$ & Khaliullin \etal \\ 
		2444595.17226  & 0.00045  & 2 &  $R$ & Khaliullin \etal \\ 
		2444595.17278  & 0.00037  & 2 &  $V$ & Khaliullin \etal \\ 
		2444595.17308  & 0.00050  & 2 &  $B$ & Khaliullin \etal \\ 
		2444988.49563  & 0.00017  & 1 &  $V$ & Khaliullin \etal \\ 
		2448499.1081~  & 0.0012~  & 1 &  $Hp$ & Hipparcos \\ 
		2457145.8527~  & 0.0013~  & 2 &  $C$ & MASCARA \\ 
		2458846.45945  & 0.00003  & 1 &  $C$ & This paper (6.02) \\ 
		2458846.45950  & 0.00003  & 1 &  $C$ & This paper (6.00) \\ 
		2458851.92656  & 0.00005  & 2 &  $C$ & This paper (6.00) \\ 
		2458851.92660  & 0.00005  & 2 &  $C$ & This paper (6.02) \\ 
		2458861.87139  & 0.00005  & 3 &  $C$ & This paper (6.02) \\ 
		2458861.87145  & 0.00005  & 3 &  $C$ & This paper (6.00) \\ 
		2458866.34985  & 0.00003  & 4 &  $C$ & This paper (6.00) \\ 
		2458866.34991  & 0.00002  & 4 &  $C$ & This paper (6.02) \\
	\end{tabular}
\end{table}

Despite the precision of the TESS data there are still some small inconsistencies. From the timings shown in Figure~\ref{fig:asymmetry} the values taken with a bright limit of $m_{TESS}=6.00$ and 6.02 have a similar time difference between the two primary and secondary eclipses, and can be further aligned by changing $\phi_2$. Figure~\ref{fig:residuals} shows the O-C diagram of these timings using the optimum value of $\phi_2$ and the optimum period. The same exercise has also been performed on Baroch \etal's timings and the four points show a similar distribution. In both cases $\phi_2$ is close to 0.45031, and similar to Khaliullin \etal's value of 0.45035, but sightly removed from Southworth's photometric solution of 0.4504. The optimum period for all the TESS solutions range from 9\fday945120 to 9\fday945168, but are very different to the long-term period near 9\fday94507108. With this period the standard deviation of the residuals in Figure~\ref{fig:residuals} increases to 0\fday00015 compared to the formal errors of typically 0\fday00005 for the KvW data measured here, or the 0\fday00003 of Baroch \etal, suggesting that for whatever reason the uncertainties of the TESS timings are seriously underestimated.

\section*{Final comments}

The movement of the primary minimum timing in Figure~\ref{fig:asymmetry} indicates that measurements based on the deepest parts of the eclipse alone can be significantly adrift from those based on more of the minimum. 
Despite their very low amplitude it seem inescapable that the pulsation variations have a significant effect on the timings of the eclipses -- in high quality data. 
These variation are too small to be  seen consistently in ground-based measurements but whether the cumulative effect on the timing could been felt is an open question. Many of the light-curves of RR~Lyn used to obtain minima have less than ideal coverage, so they probably suffer mostly from more traditional problems, but the pulsations can only increase the uncertainty and ultimately confuse searches for a third body or apsidal motion.


\end{document}